\documentclass[5p,twocolumn]{elsarticle}

\usepackage[english]{babel}

\usepackage{amsmath}
\usepackage{graphicx}   
\usepackage{subfig}
\usepackage{breqn}
\usepackage{natbib}
\usepackage{amssymb}
\usepackage{multicol,caption}
\usepackage{multicol}
\usepackage{array}
\usepackage{epstopdf}
\usepackage{color}
\usepackage[justification=justified]{caption}
\usepackage[compact]{titlesec}
\usepackage{placeins}

\titlespacing{\section}{0pt}{3ex}{2ex}
\titlespacing{\subsection}{0pt}{2ex}{2ex}
\titlespacing{\subsubsection}{0pt}{1.5ex}{1.5ex}

\newcommand{\g}{$\gamma$}
\newcommand{\hess}{H.E.S.S.}
\newcommand{\degg}{$^{\circ}$}
\newcommand{\pks}{PKS~2155-304}

\usepackage{epstopdf}
\usepackage{color}
\usepackage{dcolumn}
\usepackage{bm}
\usepackage{hyperref}
\usepackage[mathlines]{lineno}
\usepackage{upgreek} 
\usepackage{threeparttable} 
\usepackage{booktabs} 
\usepackage{siunitx}

\begin{document}

\title{Application of Deep Learning methods to analysis of Imaging Atmospheric Cherenkov Telescopes data}

\author[ecap]{I. Shilon\corref{cor1}}
\author[ecap]{M. Kraus\corref{cor2}}
\cortext[cor1]{Corresponding author;
Email: idan.shilon@fau.de}
\cortext[cor2]{Corresponding author;
Email: manuel.kraus@fau.de}
\author[ecap]{M. B\"uchele}
\author[ptd]{K.~Egberts}
\author[ecap]{T.~Fischer}
\author[hu]{T.L.~Holch}
\author[hu]{T.~Lohse}
\author[hu]{U.~Schwanke}
\author[ptd]{C.~Steppa}
\author[ecap]{S.~Funk}

\address[ecap]{Friedrich-Alexander-Universit\"at Erlangen-N\"urnberg, Erlangen Centre for Astroparticle Physics, Erwin-Rommel-Str. 1, D 91058 Erlangen, Germany}
\address[ptd]{Institut f\"ur Physik und Astronomie, Universit\"at Potsdam, Karl-Liebknecht-Str. 24/25, D 14476 Potsdam, Germany}
\address[hu]{Institut f\"ur Physik, Humboldt University of Berlin, Newtonstr. 15, D 12489 Berlin, Germany}

\date{\today}
    
\begin{abstract}

Ground based \g-ray observations with Imaging Atmospheric Cherenkov Telescopes (IACTs) play a significant role in the discovery of very high energy (E $>$ 100~GeV) \g-ray emitters. The analysis of IACT data demands a highly efficient background rejection technique, as well as methods to accurately determine the energy of the recorded \g-ray and the position of its source in the sky. We present results for background rejection and signal direction reconstruction from first studies of a novel data analysis scheme for IACT measurements. The new analysis is based on a set of Convolutional Neural Networks (CNNs) applied to images from the four H.E.S.S. phase-I telescopes. As the \hess~cameras pixels are arranged in a hexagonal array, we demonstrate two ways to use such image data to train CNNs: by resampling the images to a square grid and by applying modified convolution kernels that conserve the hexagonal grid properties.

The networks were trained on sets of Monte-Carlo simulated events and tested on both simulations and measured data from the \hess~array. A comparison between the CNN analysis to current state-of-the-art algorithms reveals a clear improvement in background rejection performance. When applied to \hess~observation data, the CNN direction reconstruction performs at a similar level as traditional methods. These results serve as a proof-of-concept for the application of CNNs to the analysis of events recorded by IACTs.

\end{abstract}


\maketitle
%


\section{\label{sec:intro}Introduction}
Since the beginning of the century, \g-ray astrophysics has been progressing at a remarkable pace. The third generation instruments of ground-based Imaging Atmospheric Cherenkov Telescopes (IACTs) have been exploring the very high energy (VHE; E $>$ 100~GeV) sky, increasing the number of known \g-ray-emitting celestial objects to more than 200~\cite{tevcat}. Current and future generations of IACTs primarily aim to investigate the origin and acceleration processes of Cosmic Rays (CRs)~\cite{cr} and identify the nature of dark matter~\cite{dm}.

The IACT technique relies on the utilization of the Earth's atmosphere as a calorimeter. When a VHE CR or \g-ray enters the atmosphere, it interacts with the nuclei in the air to initiate a cascade of particles and electromagnetic radiation, known as an Extensive Air Shower (EAS). If the primary particle is a \g-ray, it undergoes a $e^+e^-$ pair production which initiates a purely electromagnetic shower. The relativistic charged particles in the shower emit a very narrow cone of Cherenkov radiation, with an opening angle of $ \sim 1^{\circ}$, which is detectable at ground level. The small Cherenkov angle leads to a light pool with a diameter of typically 200-300\,m at ground level and a nearly uniform light density, where the integral of the intensity is correlated to the primary particle's energy. Electromagnetic showers are characterized by an elliptically shaped shower image. If the primary particle is a CR, a hadronic shower develops. Although such hadronic showers often have electromagnetic sub-shower components as well, they lead to a typically more irregular shape of the image.

IACTs are able to detect and record images of the Cherenkov light emitted by the secondary particles in the EAS. Such images generally allow one to gather sufficient information to separate the \g-ray signal from the dominant CR background and reconstruct the source position and energy of the primary \g-ray. To record a sufficiently accurate shower image, an IACT camera must be able to capture the very brief Cherenkov light flash, which lasts for a few nanoseconds. In addition, the optical Point-Spread Function (PSF) and camera pixel size should ideally be smaller than the angular dimension of the \g-ray shower. Increasing the IACT camera resolution enables a more accurate computation of the shower axis, which has an intrinsic transverse angular size of only a few arc-minutes. Nevertheless, stochastic fluctuations in the shower development impose a limiting factor on the performance of an IACT.

The analysis of EAS images improves significantly when observing the showers from several angles~\cite{stereo}. A current generation IACT system that utilizes such stereoscopic analysis of EAS images is the High Energy Stereoscopic System (H.E.S.S.)~\cite{hess_1}. The \hess~array is located in the Khomas highland in Namibia (23$^{\circ}$16'18''~S, 16$^{\circ}$30'01''~E). It consists of four 12\,m diameter Cherenkov Telescopes (CT1-4), built between 2002 and 2004, and a fifth, 28\,m diameter telescope (CT5), built in 2012. The CT1-4 telescopes have a total field of view (FoV) of 5\degg and an energy threshold of about 100\,GeV. Thanks to its large mirror surface, the minimum \g-ray energy that CT5 can trigger on is $\sim$ 30~GeV, with an FoV of 3.5\degg. Each of the small telescopes is equipped with a camera containing 960 photo-multipliers (PMTs), while the camera of CT5 contains 2048 PMTs.

Analysis of IACT images relies on the extraction of relevant features from the camera pixel data. Whether those features are a vector of parameters representing the image, such as the image moments, or the full photo-electron intensity count in each pixel, in order to detect and study VHE \g-ray sources an IACT analysis method must be able to perform each of the following three tasks:

\begin{enumerate}
\item \emph{Background rejection}: separate the \g-ray induced signal from the much more prevalent background of hadron-induced showers, through identification of shape features in the image. 
\item \emph{Direction reconstruction}: reconstruct the position of the origin of those events classified as signal, through calculation of the shower image axis. Observation of the EAS with a stereoscopic system significantly improves the direction reconstruction resolution.
\item \emph{Energy reconstruction}: reconstruct the primary particle's energy for those events classified as signal, through the total image intensity and the shower impact point on the ground, relative to the telescopes.
\end{enumerate}

H.E.S.S. currently applies three main reconstruction techniques in its analysis chain. The first relies solely on the so-called Hillas parameters~\cite{hillas}, the image moments derived from the distribution of the measured intensities in the individual camera pixels, calculated after an image cleaning step~\cite{hess_crab}. Given the approximately elliptical shape of typical signal camera images, the arrival direction of an EAS event is reconstructed by tracing the major axis of the image, which corresponds to the projected direction of the shower in the camera FoV, to the \g-ray source in the sky. For stereoscopic observations, the major axes of EAS images from the participating telescopes are calculated in a common camera reference frame and the intersection points of all axes pairs are found. A weighted average, based on image amplitude and the angle between the axes, is then taken of all intersection points to provide an estimate of the arrival direction of the primary \g-ray \cite{hess_crab}. A similar procedure, involving the intersections of the directions between the image centroid and the optical axis, is then performed in a common plane perpendicular to the pointing direction, to determine the shower impact point on the ground.

The two other techniques utilize likelihood fitting of camera pixel amplitudes to semi-analytical shower models (the Model++ method~\cite{model}) or template libraries from Monte-Carlo (MC) simulations (the ImPACT method~\cite{impact}). A maximum likelihood fit is performed to find the best-fit shower parameters. These analysis methods show significantly better performance with respect to the Hillas analysis, particularly for direction reconstruction. 

The Hillas and ImPACT reconstruction methods rely on a Boosted Decision Tree (BDT)~\cite{hess_bdt} for the background rejection stage. The BDT uses a set of parameters, derived from Hillas-based event reconstruction, to classify events. These parameters include a comparison of the image width and length to the expected mean values for both \g-ray and hadron induced showers, the average spread in energy reconstruction between the triggered telescopes and the reconstructed height of maximum of the air shower. The BDTs are trained in a set of energy and zenith angle bins. The Model++ scheme uses a scaled shower goodness-of-fit parameter to separate signal from background.

The analysis of IACT data obviously relies on the correct extraction of relevant features from the EAS images. Huge improvements over the last decade in computational power, particularly in the usage of GPUs for matrix operations, suggest that more computationally demanding algorithms could be utilized to boost the performance of such analysis chains. Specifically, Deep Learning (DL) techniques for object recognition in images and sequences of images, where a machine learns relevant features from the entire image matrix, are a clear candidate for such a family of algorithms. In general, DL concerns the application of complex Artificial Neural Networks (ANNs) to hierarchical learning tasks. For computer vision, Convolutional Neural Networks (CNNs) were designed and developed specifically to perform image recognition tasks. In this work we demonstrate how the application of DL techniques, relying on CNNs for recognition of features in EAS images, to H.E.S.S. data enhances the analysis of astrophysical point-sources observed by the H.E.S.S. CT1-4 telescopes.

In the following section, we provide a short description of the DL algorithms which we have used throughout this work. Section \ref{sec:ds} provides details regarding the data-sets we created to train and test our networks. In section \ref{sec:prep} we describe the data pre-processing approaches we have taken in order to feed H.E.S.S. data into a DL framework. Sections \ref{sec:clas} and \ref{sec:dir} describe the training process and provide test results for classification and direction reconstruction, respectively. Next, in section \ref{sec:real}, we apply our DL models to real data to present our analysis results and compare them to current \hess~analysis methods, by utilizing the \hess~Analysis Package (HAP). We conclude our findings in section~\ref{sec:sum}.

\section{Deep Learning Methods for IACT Data}
\label{sec:dl}

In its essence, DL is based on the use of ANNs. The basic unit of an ANN is the multi-layer perceptron (MLP), that can be viewed as a weighted directed graph in which the "neurons" are graph nodes and directed edges with weights are connections between input and output. As the name suggests, an MLP graph contains a number of layers, starting with an input layer where each node receives all input variables~$x_i$. The number of nodes in the output layer $y_k$ is the number of desired output labels in the case of a regression model or the number of classes in a classification model. In between the input and output layers, additional layers are introduced, where the number of nodes in each layer and the number of layers are free hyper-parameters. These graph layers, commonly known as hidden layers, allow the introduction of non-linearity to the model by means of a non-linear activation function. In the context of DL, such layers are usually referred to as fully connected (FC) or dense layers.

All the MLP neurons, but those in the input layer, receive their input from each of the nodes in the preceding layer. The output from each neuron in the hidden layer is given by $f(\vec{w} \cdot \vec{x} + b)$, where $f$ is called the activation function, $w_j$ are the weights and $b$ is an additional degree of freedom, called the bias. Typical activation functions for DL networks are the hyperbolic tangent function and the Rectified Linear Unit (ReLU), defined by $ReLU(x)~=~\max(0, x)$~\cite{relu}. For the output layer $f$ is the identity function. The learning process is accomplished by the 'backward propagation of errors' (or BackProp) algorithm~\cite{backprop}. The errors are calculated by means of a loss function, predefined in accordance with the learning task.

In a classical machine learning setting, one would ``engineer" the features in the data-set in order to improve the performance of the ANN. As explained in the previous section, we wish to apply CNN based networks to \hess~data. CNNs are a specialized kind of a neural network for processing data that has a known, grid-like, structure \cite{dl_cnn}. In the 2D case, CNNs take a complete 2D grid of image pixels as input. Another desirable characteristic of CNNs is the fact that they automatically induce feature engineering of the input, meaning that CNNs learn to identify the relevant and important features in the training data in order to optimize their performance for a certain task.

A typical architecture of a CNN includes numerous Convolutional Layers (CLs), followed by a number of dense layers (usually $\leq 3$). A CL typically comprises three stages: a convolution stage, an activation stage and a pooling stage. It should be noted that variations of this structure are common and in fact we make use of a different layer structure for our regression tasks (see Sec.~\ref{sec:dir}).

The convolution stage relies on a set of learnable filters with a fixed size, which is spatially smaller than the 2D input. Each of the filters is convolved across the entire width and height dimensions of the 2D image matrix to produce linear outputs. The small spatial size of the filters allows the CNN to detect meaningful features (e.g. edges of a shape) which occupy only a small portion of the image. The convolution operation involves an element-wise multiplication between the filter and local patches of the image of the same size as the kernel. This {\it local} (or sparse) connectivity significantly reduces the number of free parameters and memory requirements of the model. Moreover, each parameter of the filter is ``shared" across the image, in the sense that it is applied at every position of the image (disregarding boundary effects). {\it Parameter sharing} through patches of the image introduces {\it translational equivariance}, meaning that a shift in the input leads to the same shift in the output. 

The convolution stage of a CL is followed by an activation stage, where each linear activation is fed into a nonlinear activation function. In the pooling stage, a pooling function~\cite{pool} replaces the output of the layer at a certain location with a summary statistic of the nearby outputs, such as the maximum output in the neighbourhood of that location. The pooling operation makes the output of the convolution become approximately invariant to small translations in the input.

In many industry applications, CNNs are required to deal with coloured images. In such cases, the images are represented by a 3D grid, where the three layers along the depth dimension represent the RGB color components of each pixel. The convolution is done with 3D filters to convolve along the RGB layers, called channels in DL terminology. For IACT images, we may treat each of the telescope images as a single channel of the event image. In this case the depth is defined by the maximum number of participating telescopes $t$ and the filters have dimensions of $m \times n \times t$, where $m$ and $n$ are the width and length of the filter, respectively, and in our case $2 \leq t \leq 4$. This procedure will be further explained below.

In addition to the telescope channel representation, one may view the images of an EAS event as a temporal sequence of images. The ordering of the sequence can be determined by the time order of the triggering telescopes. With such event representation, one may apply a Recurrent Neural Network (RNN)~\cite{dl_cnn}, where for example the output of the CLs are fed into a recurrent cell before it is sent to the dense layers. RNNs enable machines to be persistent by finding correlations between the different inputs in the sequence. This implies that an RNN has a ``memory" that captures information from previously analysed data in the sequence. It should be noted, however, that the temporal correlations learned by the recurrent cell can be bi-directional as well (i.e. looking at both past and future data during the learning process). However, the computation requirements for such case are heavier and we have not found sufficiently convincing reasoning for applying it here. The recurrent cell applied in this work is the so-called Long Short Term Memory (LSTM) cell~\cite{lstm}.

To implement our models, incorporating the algorithms described above, we have been utilizing the TensorFlow~\cite{tf} DL framework. The models were trained on a machine with two Nvidia\textsuperscript{TM} GeForce GTX 1080 GPUs. We took the data parallelization approach to accelerate the training process. 

\section{Training and Test Data-sets}
\label{sec:ds}

The parameter distributions that describe a data-set used to train a neural network are incorporated in the prior probability of the resulting classifier or regressor. This statement is clear if one considers the network as a machine that predicts the function that is likely to have generated the data. Therefore, one must carefully consider the distributions of parameters that describe the training data-set before initiating the training process. For example, for a classification task, a training set should contain an equal number of examples from each class. In the case of IACT data, and in particular for regression tasks, one might consider training on sets with different energy spectra or offset distributions to help the learner converge towards the desired predictor. In addition, it is a common rule of thumb that deep networks perform better with larger training data-sets.

\subsection{Training data-sets}

Throughout this work, all data-sets that were used for training the different networks comprise events generated by Monte-Carlo (MC) simulations. These events are obtained by simulating the interaction of \g-rays (i.e. signal) and protons (i.e. background) with the atmosphere using the CORSIKA software \cite{cors1}. Following the shower simulation, the response of the H.E.S.S. telescopes is simulated using the sim\_telarray package \cite{corsika} in order to generate the telescope images.

The three analysis goals described in Sec. \ref{sec:intro} suggest two basic types of networks. The background rejection goal calls for a classification network, where an event is classified into one of two possible groups (i.e. signal and background). The two reconstruction goals are regression tasks, where the network is trained to predict a continuous parameter based on the image input. To that end, we chose two data-sets, one for classification training and another for regression training. 

The classification training-set includes 2$\times 10^6$ events, where the ratio of signal events to background events is one. The equal number of signal and background events in this data-set implies that the data-set is balanced in terms of the class-labels. For the regression tasks, we use a training-set comprised of 1$\times 10^6$ MC \g-events. All events were simulated as diffuse emission around 20$^{\circ}$ zenith and 180$^{\circ}$ azimuth and powerlaw spectral index of $-2$.

The differences between the two data-sets are in the simulated view-cone (i.e. the solid-angle around the telescopes pointing position, within which showers are generated), the energy ranges and the telescopes' optical efficiency. For the classification set both particle types are simulated with a view-cone of 2.5$^{\circ}$ and the signal events have energies from 20\,GeV to 100\,TeV while background events are simulated with energies that range from 100\,GeV to 200\,TeV. The events in the regression data-set are simulated with a view-cone of 3$^{\circ}$ and have energies between 5\,GeV and 100\,TeV. The different parameter ranges come from our choice to rely on existing \hess~simulations in order to save valuable computation time.

Because we standardize our images as part of the pre-processing stage (see Sec. \ref{sec:prep}), the CNNs are blind to the optical efficiency of telescopes. Nevertheless, for reference we state that the classification sets are created from the so called \hess~phase1 simulations and the regression sets contain \hess~phase2b5 simulations, where the phases refer to a state of the \hess~array.

The raw simulated images are cleaned according to the standard H.E.S.S. cleaning scheme \cite{hess_crab}. To take advantage of the stereoscopic observations of \hess, our simulation sets consists of events that survived the image cleaning in at least two of the CT1-4 telescopes (referred to as multiplicity-2). CT5 images are omitted from all data-sets. 

We would like to point out that the training set numbers given above are without the usage of pre-selection cuts. This means, particularly due to the large energy range of the data-set, that many of the events ($\sim$ 30\% on average) in the training data are truncated and would not have passed the standard pre-selection cuts of the HAP~chain (see Sec. \ref{sec:bench}). When applying pre-selection cuts to the training data, we observed enhanced performance solely for the direction reconstruction on simulated events. In this case, the number of events in the training set is 620\,k.

In addition, the training sets were not binned in any parameter (other than the zenith angle). Taking the binned training approach could, in principle, increase the accuracy score of the classifier. However, this means that when coming to analyse real data each event would have to be sent to its corresponding classifier. This requires knowledge about the particle properties (e.g. the energy for energy binned training) prior to the reconstruction stage. As the energy and direction of the \g-ray are not necessarily known at the classification stage, this approach was not favoured at this proof-of-concept stage of the project.

\subsection{Benchmark test data-sets}
\label{sec:bench}

In addition to the training sets, we have also created independent test data-sets in order to serve as benchmark and test the performance of the classifiers and regressors in a statistically significant way. The benchmark sets are used to demonstrate how well a classifier or a regressor generalizes to an arbitrary set of events from the simulation data - excluding events that were used in the training process. Therefore, the benchmark test data-sets are sub-sets of the relevant event distributions from the data-set from which the training sets were created. One should note that the benchmark sets did not serve as validation sets, i.e. these data-sets were not used to tune network hyper-parameters. The benchmark test results were obtained only after the work on a classifier or a regressor was completed. 

Another purpose of creating benchmark sets is to make a comparison between the DL based results and the Hillas and ImPACT analyses of HAP. These analysis schemes are typically not able to correctly classify or reconstruct events that do not pass a set of defined pre-selection cuts, while in some circumstances a DL based analysis is able to do so. For the HAP analyses, cuts on the minimum image total amplitude (denoted as the size parameter) and the maximum distance between the Hillas ellipse centre-of-gravity and the camera centre (denoted as the local-distance parameter) are applied. The local-distance cut is used to reduce effects of image truncation at the edge of the camera. We have used a standard set of pre-selection cuts, where the minimum size parameter of an image is 60\,p.e. and the maximum local distance of an image is 0.525~m  (equivalent to 2\degg~in the camera FoV). Together with the multiplicity cut, the pre-selection cuts mean that each surviving event must have at least two telescope images that survive the pre-selection cuts. 

For the classification task, the size of the free benchmark set (i.e. without pre-selection cuts) is 2$\times$320\,k events, while the pre-selected benchmark set contains 2$\times$196\,k events. For the regression tasks, the benchmark set contains $\sim$500\,k \g-events that pass the pre-selection cuts from a point-like source with an offset of 0.5\degg to the pointing position. The test results on simulated events we present were obtained using these benchmark test sets.

\section{Image Pre-processing}
\label{sec:prep}

\begin{figure*}
	\centering
	\includegraphics[width=0.8\textwidth]{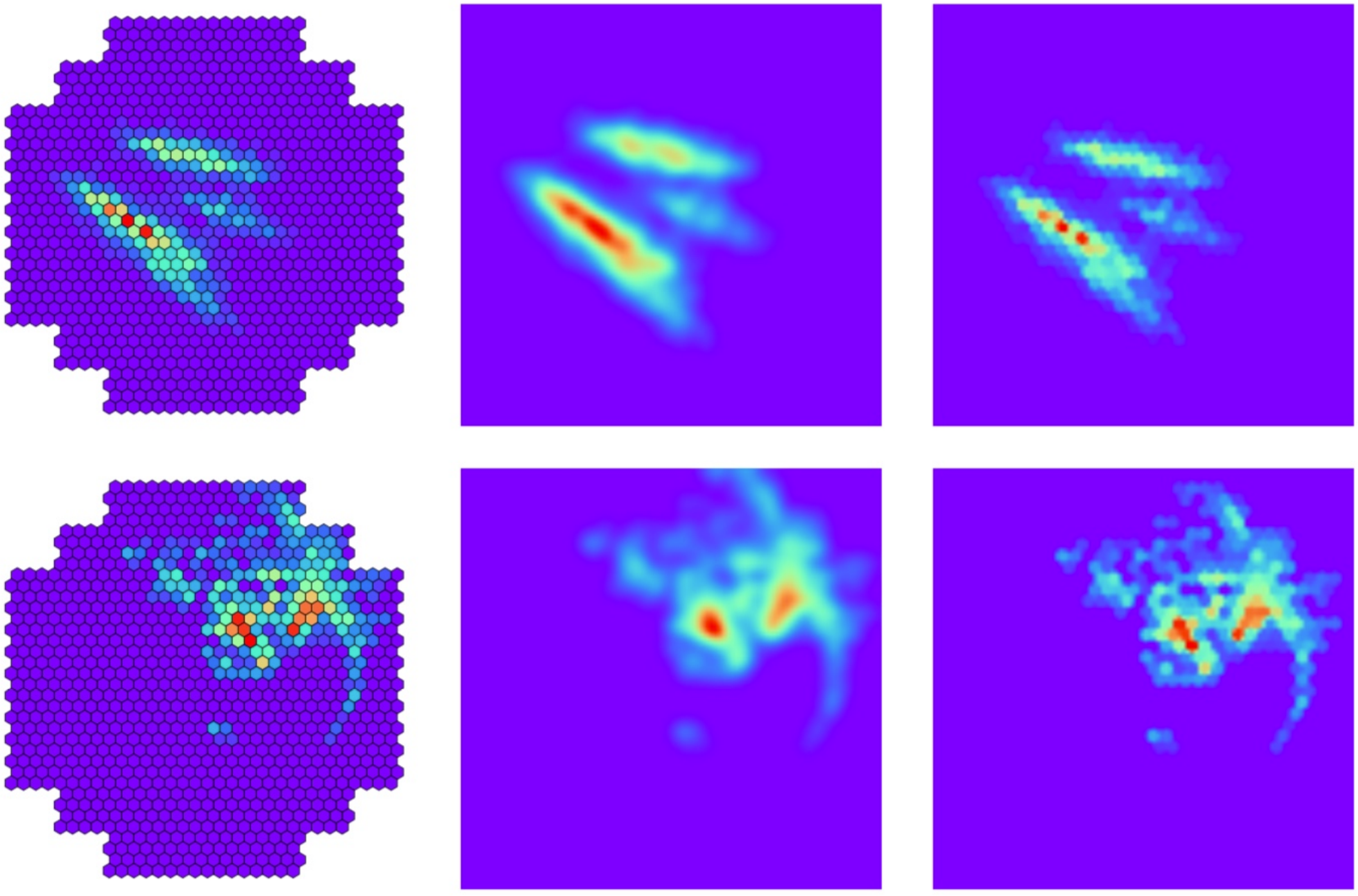}
	\caption{Telescope images of a \g-ray (\textit{top row}) and a hadronic (\textit{bottom row}) event. The images consist of numerous event images laid in a common camera plane, for visualization. \textit{Left column}: Original camera pixel intensities. The square images show the same event sampled with Gaussian smoothing (\textit{left}) and by rebinning an hexagonal histogram (\textit{right}). The square images are oversampled with a resolution of 100 $\times$ 100.}
	\label{fig_resamp}
\end{figure*}

The photomultipliers (PMTs) tubes in the cameras of the H.E.S.S. telescopes are arranged in a hexagonal grid. However, CNN implementations in standard DL frameworks are designed to process arrays of data points arranged in square grids. To process H.E.S.S. data with these frameworks, it is therefore necessary to pre-process the camera images and map the image data onto a square grid. Converting data points from a hexagonal to a square grid is not trivial, as the two lattices respect different symmetries (6-fold symmetry for the hexagonal lattice and 4-fold symmetry for the square lattice). The challenge is therefore to perform discrete convolution operations on pre-processed telescope data using the methods provided by DL frameworks, while conserving the spatial symmetries and intensity distributions of the original data as much as possible.

There are in principle two possibilities to address this complication. The conservative approach is based on the resampling of the hexagonal data. The idea is to transform the data itself in such a way that the original image properties are approximately translated to a square grid. An alternative is the construction of custom convolution kernels that conserve the properties of the hexagonal grid. The data points are therefore rearranged into a square grid, to which the custom convolution kernels are applied.

While the resampling approach can only approximately conserve the hexagonal image properties (where the degree of distortion depends on the specific method and resampling resolution), it allows one to take advantage of the full set of functionalities and methods provided by modern DL framework and apply state-of-the-art CNN architectures. This is not the case for the custom hexagonal kernels, as such task specific operations require manual adaptation of the common and available methods. To explore the applicability of such costum kernels to \hess~data, an extension to the PyTorch DL framework \cite{pytorch} has been developed \cite{ptrepo}, providing flexible implementations of convolution and pooling operations for hexagonally sampled input data. This approach shows promising initial results on simulated data, but is still being developed to be applied to real-world data. We present here the principles of the technique and plan to show applications of this algorithm in a future publication.
	
\subsection{Comparison of resampling methods}
\label{sec_sampling}

There are several common approaches to generate square images from hexagonal data. For example, viewing the camera intensity map as a grid of points, each of which located at the coordinates of the pixel centre, holding the pixel intensity, opens up a multitude of interpolation methods such as linear and cubic interpolation. Interpolation methods are probably the most widely used approach for image processing. To interpolate, pixel values are interpreted as discrete samples of a continuous function - the image intensity distribution in the case of \hess. Constructing such a function allows resampling the original image at an arbitrary set of points. The interpolated values obey some relationship to neighbouring original image values depending on the choice of interpolation method. For the \hess~images, problematic behaviour arises in the corners of the cameras because of the large distance between square and hexagonal pixel centres. This can be circumvented by introducing artificial zero valued pixels in order to mask the camera corners. One should note that linear and cubic interpolation do not preserve the total image intensity.
	
On the other hand, interpreting the camera's PMTs as bins in an hexagonal histogram that collect single photons is more realistic than a centre point approximation and the histogram can be efficiently rebinned into a square histogram. Camera pixel values are related to the number of photons collected by the corresponding PMT's photocathode. Interpreting the image as a histogram is thus more physical than concentrating the pixel value to a single point. For our purposes we would like to know what an image taken with a camera consisting of a square grid of PMTs would look like. This can be achieved by rebinning the hexagonal histogram into a square one. This allows using an arbitrary resolution and implies the conservation of total image intensity. Rebinning can be formulated as a single sparse matrix-vector multiplication, yielding low computation times. For further details on the rebinning method applied here, see~\cite{icrc}.

In \cite{icrc} we also show the results of a thorough comparison between interpolation, rebinning, smoothing and oversampling methods, where smoothing refers to methods such as Gaussian smoothed sampling and an example of oversampling IACT images can be found in \cite{ver}. The comparison investigated the influence of the resampling methods on artificial telescope images, generated by a 2D Gaussian function that was sampled to a camera-like hexagonal grid. The results of this study show that cubic interpolation excels at shape conservation. However, it is by far the most computationally expensive method. Linear interpolation and rebinning exhibit similar performance that is reasonably close to cubic interpolation. Nevertheless, the rebinning is somewhat more computationally efficient relative to linear interpolation. Smoothing and oversampling methods were ruled out as resampling method for \hess~images in the study. Fig.~\ref{fig_resamp} shows examples of pre-processed images, resampled with a Gaussian filter and our rebinning method.

\subsection{Resampling of \hess~images}
\label{iact_sampling}

Following our comparison study and to study different approaches, we generate resampled image inputs using both rebinning and linear interpolation methods. The rebinned images were used for classification models, where data-sets are generally large, and the interpolated images were used for direction reconstruction models. The resolution of the resampled images significantly affects the performance of the analysis, where we observed that too high or too low resolutions degrade performance. In both cases, images were resampled with a resolution of 64$\,\times\,$64 pixels (due to the cameras' aspect ratio the true resampling resolution is 64$\,\times\,$62; To get a square image we pad the images appropriately). This leads to a ratio of roughly four between the number of resampled pixels and camera pixels.

As part of the pre-processing stage, we standardize each image, so that it has a mean intensity of zero and a standard deviation of one. Besides accelerating convergence, the standardization of the images effectively makes the networks invariant to different telescope optical efficiencies. This means that the classification and direction reconstruction are relying solely on shape features and thus can be applied to analysis of real data without accounting for the telescope optical efficiency which degrades over time. For energy reconstruction, one may reconstruct the shower impact point on the ground using a CNN, thereby conserving the optical efficiency invariance. We discuss possible implementations of a DL-based energy reconstruction in more detail in Sec.~\ref{sec:sum}.

\subsection{Hexagonal Convolutions}
\label{sec_hexconv}

\begin{figure*}
\centering
\includegraphics[width=0.8\textwidth]{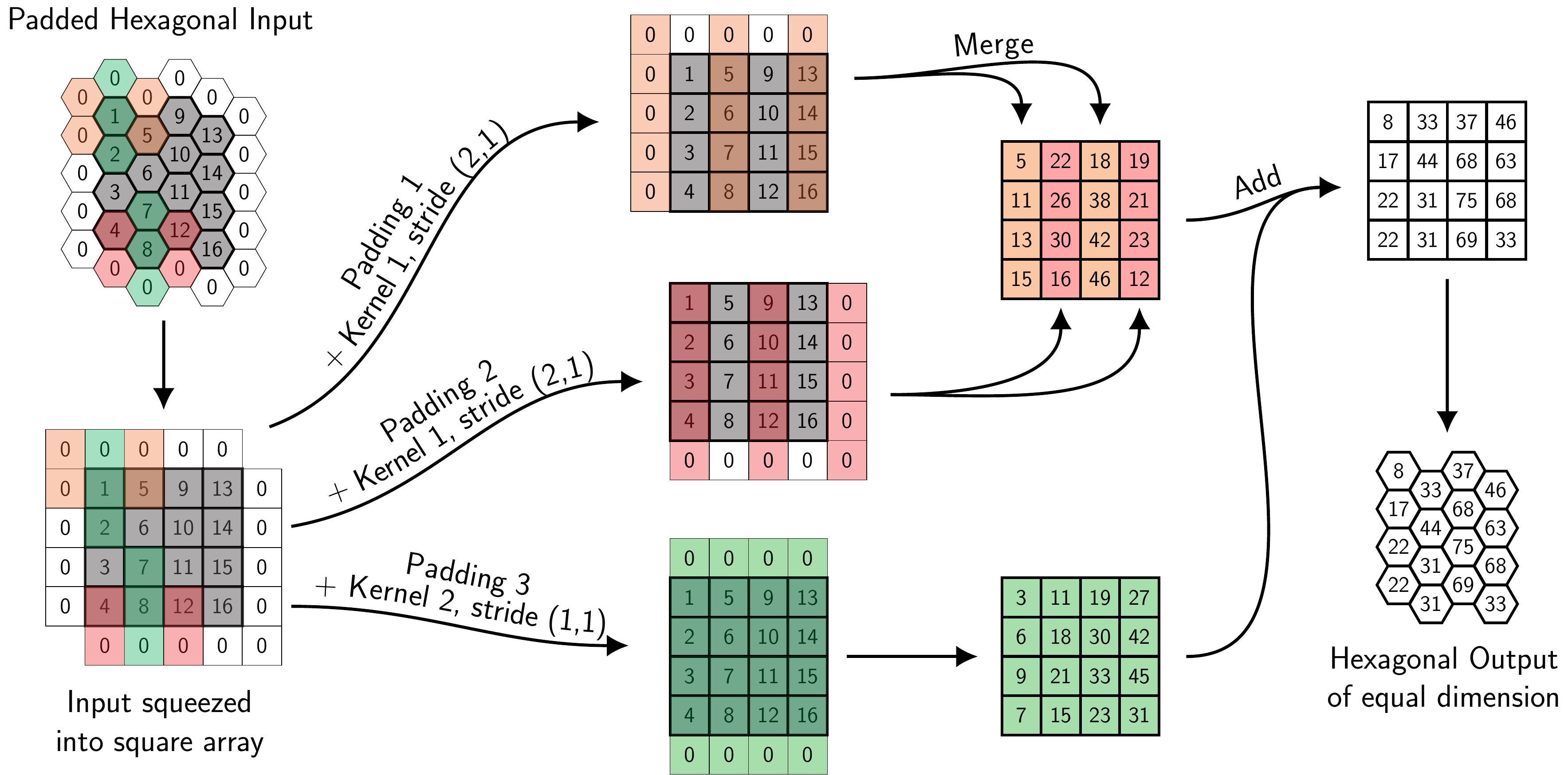}
\caption{Schematic description of the individual steps of a hexagonal next-neighbour convolution. All kernel weights are set to~1. The input image consists of $4 \times 4$ hexagonal toy pixel values with ascending integer values and is padded with one layer of zeros in order to preserve the dimension of the input. In the first step, the hexagonal image is squeezed into a square layout. Then, the hexagonal kernel is split into two square sub-kernels: Kernel~1 (red \& orange, $2 \times 2$, dilation $(2,1)$, stride $(2,1)$) and kernel~2 (green, $1 \times 3$, stride $(1,1)$). Kernel~1 must be applied separately to even and odd columns of the squeezed image. The output of these two operations has to be interchangeably merged to preserve the spatial relationship of the columns. In the last step, the merged array is summed with the result of the squeezed image convolution with kernel 2. The resulting square array is equivalent to the result of a hexagonal convolution map.}
\label{fig_hex_conv_example}
\end{figure*}

The basis to design custom hexagonal kernels, while relying on standard image processing algorithms that assume the input to be sampled on a square grid, is the rearrangement of the hexagonally sampled data and the hexagonal kernels. To that end, our implementation of the custom kernels assumes that the original input is \textit{squeezed} into a square layout. Sub-kernels are then convolved with specific patches of the squeezed image according to the hexagonal layout. Squeezing means that the hexagonal lattice is zero-padded (according to the desired size of the hexagonal convolution kernels) and that the hexagonal lattice cells are interpreted as square cells. To complete the squeezing, each protruding lattice column is shifted by $1/2$ of the vertical lattice spacing to get a well defined square grid. The horizontal lattice spacing is kept fixed. This process, depicted in the first step of Fig.~\ref{fig_hex_conv_example}, results in a square grid with equal length columns.

Next, the hexagonally-symmetric convolution is accounted for by splitting the hexagonal convolution kernel into hexagonal sub-kernels. These sub-kernels need to be re-defined on the squeezed grid in order to perform dilated convolutions using standard methods. The shape, dilation and stride of each of the re-defined sub-kernels are chosen so that they are effectively applied only to pixels that are the hexagonal $n$th neighbours of the centre pixel of the corresponding hexagonal image patch, where $n$ is a positive integer. Combining the sub-kernels' convolution outputs in an appropriate way yields a map that is equivalent to the result of applying a hexagonal kernel to the centre pixel of the image patch.

The detailed description of defining the custom sub-kernels is out of the scope of this paper, but is presented in \cite{ptrepo}. Here we merely supply the reader with an example, shown in Fig.~\ref{fig_hex_conv_example}. In the figure, a toy input tensor is convolved with a hexagonal next-neighbour (NN) convolution kernel (i.e. where the kernel covers all direct neighbour cells of the centering cell in the hexagonal layout) and the individual steps for squeezing a hexagonal tensor and assigning the custom kernels to the squeezed image are schematically shown.

\section{Background Rejection Classifier}
\label{sec:clas}
Compared to satellite-based detectors IACTs provide large effective detector areas. However, the vast majority of events recorded by an IACT contain hadronic CR background. The ability to correctly reject background events without loosing signal events is therefore a key aspect that determines the sensitivity of an IACT and hence serves as a primary goal for this work.

During this study we have learned that CNNs trained on simulated events exhibit different performance when tested on a MC test-set and when analysing real-data. This statement holds for all three analysis tasks. For classification, the MC/real-data discrepancy manifests itself in the following way. When compared to the real data performance of the HAP BDT classifier, we found that classifiers based on a standard CNN architecture tend to misclassify events that triggered three or four of the telescopes, although these classifiers were showing the best performance on the benchmark sets. However, by combining a CNN with an RNN this mismatch is considerably relaxed and the real world performance of the classifier improves significantly. Therefore, we present here the results obtained with the latter, denoted by CRNN. This interim solution suppresses the discrepancy effects, but is certainly not an optimal and robust way to address it. The discrepancy between simulation and observation has important potential implications for analysis of real data and we address them further in Sec.~\ref{sec:sum}.

In the CRNN model we treat each telescope image as part of a sequence ordered by the size parameter of each image. This ordering assumes that EAS images with higher size parameter are generated first in telescopes that are closer to the impact point of the shower axis on the ground and is expected to compensate for the lack of temporal data in the event images data.

\begin{figure*}
\centering
	\subfloat[ROC curves with matching AUC values. \label{fig:roc_mc}]{\includegraphics[width=.5\textwidth]{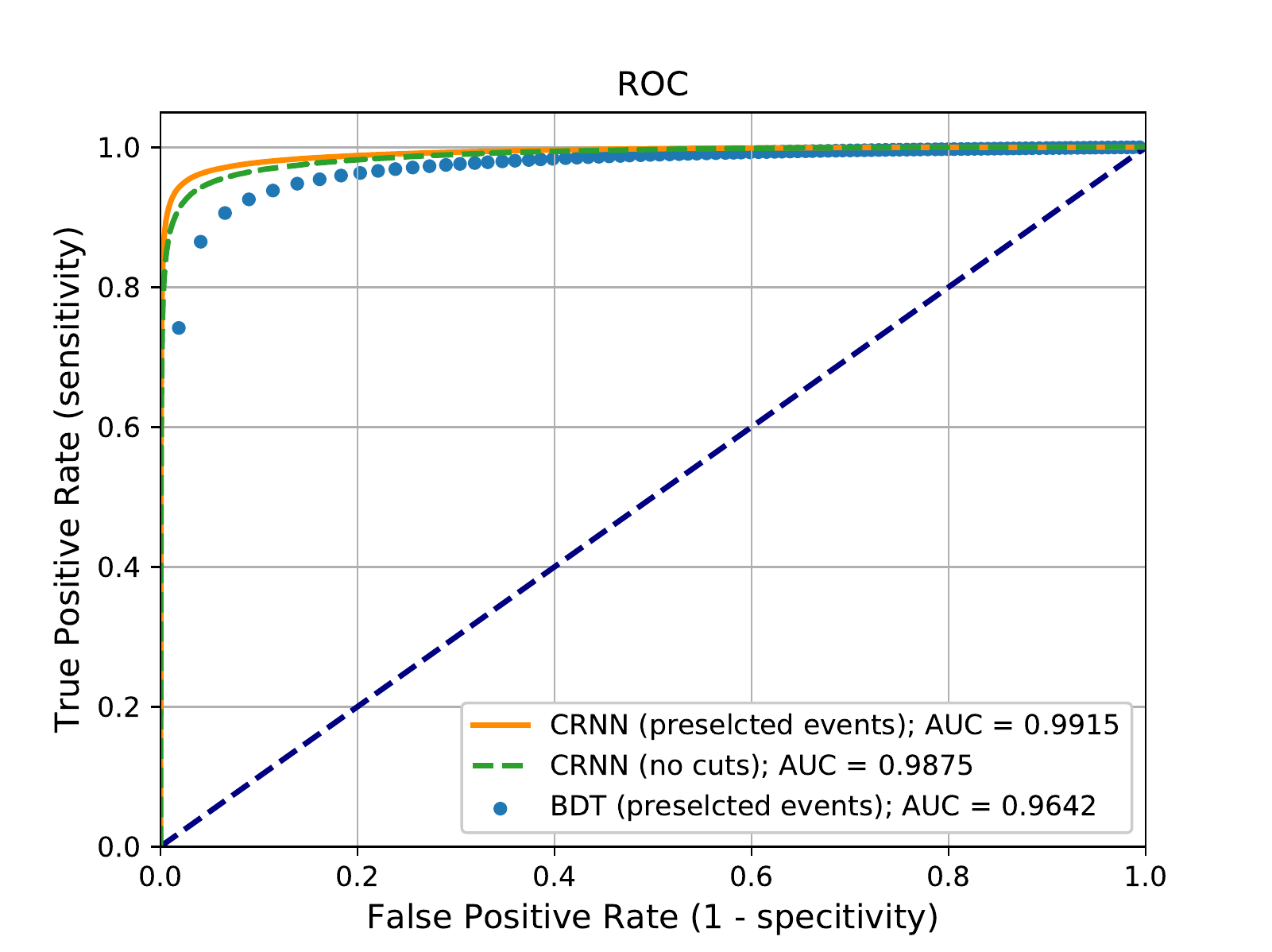}
	}
	\subfloat[CRNN $\zeta$-score distribution for simulated signal and background events. \label{fig:z_withcuts}]{\includegraphics[width=.5\textwidth]{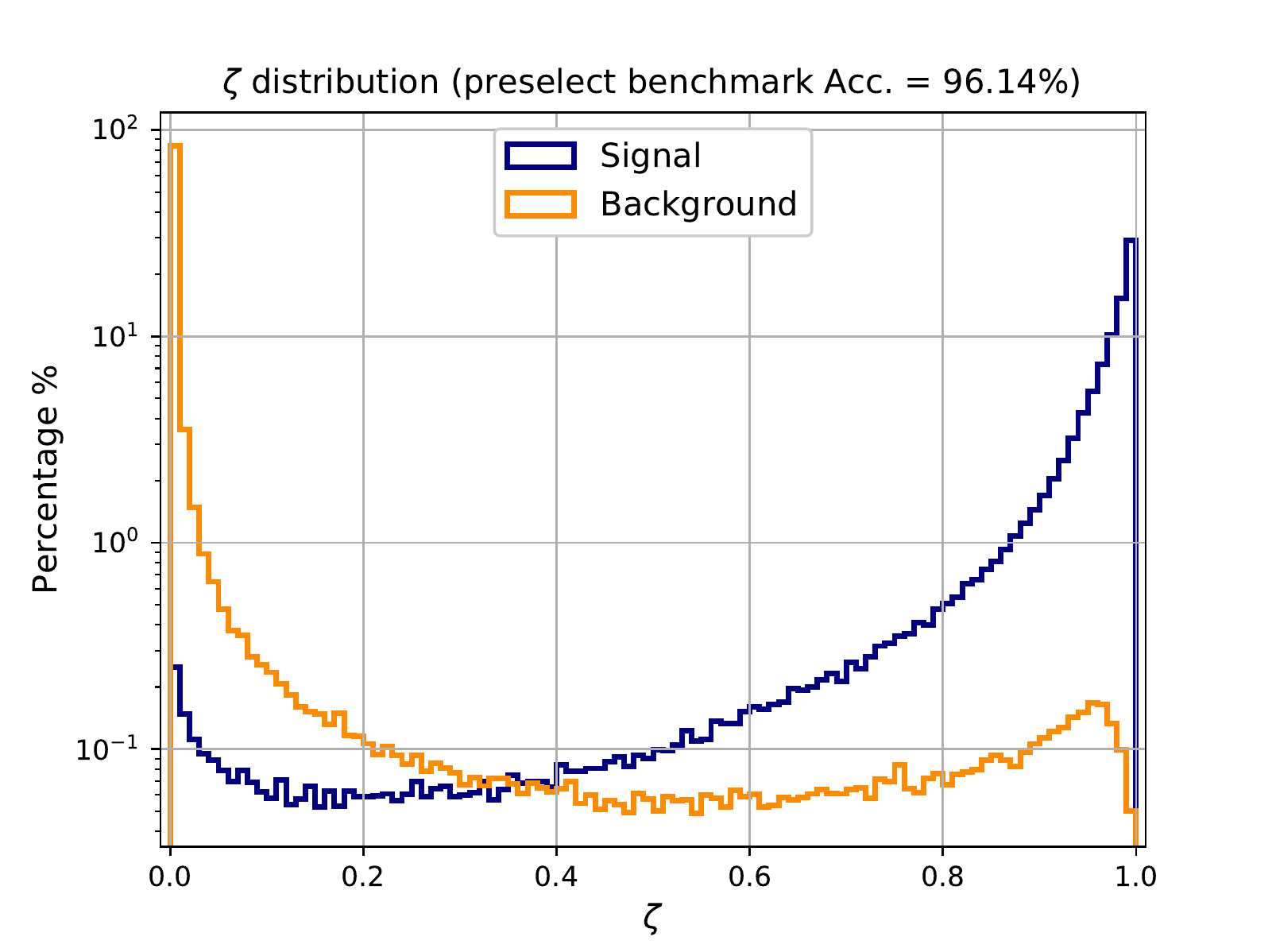}
	}
\caption{ROC curves for the CRNN classifier and the \hess~BDT classifier (left) and the $\zeta$-score distribution for the CRNN classifier, obtained on the benchmark data-set with pre-selection cuts.}
\label{fig:clas}
\end{figure*}

The CRNN architecture is based on the idea that a series of CLs can be used to find a vector representation of the 2D event images. Then, by ordering these vectors according to a size-based sequence, an RNN cell can discover correlations between the vector representation of the different event images. Lastly, the RNN cell output is fed into a dense network.

To implement this idea, we used a network containing three CLs for the image representation stage. A 0.5 dropout rate (see \cite{dropout}) was applied to the vector outputs of the last CL. These vectors are then fed into an LSTM cell, according to the image size order. The output of the recurrent cell goes through another dropout node and is then fed into a two-layer dense network with dropout after each layer. To further reduce overfitting, we regularize each dense layer by applying weight decay with a constant of 0.004. The last layer of the network is a simple linear layer with two nodes. To classify inputs, one feeds the linear inputs to a softmax function to yield a probabilistic measure and predict the class of the input image.

To calculate our test accuracy, we used a $\zeta$ threshold of 0.5, where $\zeta$ denotes the signal class softmax value for a single event. An event is classified as a \g-ray when it receives a $\zeta$-score $>$ 0.5, otherwise it is labeled as background ($\zeta$ can be thus interpreted as the ``probability to be a \g-ray"). For this threshold, our network achieves a test set total accuracy of 95.4\% on the base benchmark set and 96.1\% on the pre-selected benchmark set (where total accuracy accounts for both \g-rays and protons which are correctly classified). We illustrate the general performance of the classifier by the ROC curve and the derived area-under-curve (AUC) metric. The test results on the classification benchmark sets are shown in Fig. \ref{fig:clas}. Fig. \ref{fig:roc_mc} shows the ROC curve, along with corresponding AUC values, of the CRNN classifier when applied to both classification benchmark sets, i.e. with and without pre-selection cuts, as well as the ROC curve of the standard \hess~BDT when classifying only the pre-selected events. Fig. \ref{fig:z_withcuts} shows the CRNN $\zeta$-score distribution of the events that pass pre-selection~cuts. 

Fig.  \ref{fig:roc_mc} demonstrates that, unlike the BDT classifier, the CRNN classifier is quite robust against simulated images that are not fully contained in the camera. A BDT classification relies on the Hillas parameters, which may be significantly biased when a shower image is truncated. The CRNN, however, searches for common shape features in patches of the image and hence is less sensitive to distortions originated by truncation or broken pixels.

\section{Direction Reconstruction Regressor}
\label{sec:dir}
 
In contrast to the geometrical direction reconstruction that was described in Sec. \ref{sec:intro}, a neural network learns to predict the shower direction based on the features the convolutional filters extract from the images. We have seen that for regression, a CNN with a channel representation of event images generally outperforms architectures that include a recurrent cell.

For the direction reconstruction task, we incorporate convolutional layers with a slightly different structure than that presented in Sec. \ref{sec:dl}. Here, the structure of the convolution layers comprise one convolution stage, one nonlinear stage, a second convolution stage, a second nonlinear stage and only then a pooling stage. We denote this layer structure by 2-1-CL. Multiple stacked convolution with nonlinear stages can develop more complex features of the input volume, before the pooling stage down-samples the non-linear convolution output. This is generally a good idea for larger and deeper networks, that are necessary for the regression task. The need for a deeper network can be understood by viewing the continuous labels of a regression network as an infinite set of discrete classes. 

The network architecture for the interpolated images input includes five 2-1-CLs followed by four dense layers. We use the same weight decay as in Sec. \ref{sec:clas} for the dense layers and apply dropout after each dense layer with a rate of 0.8. The loss is calculated with the L1 distance to control the generic differences between features of low and high energy events, where the latter are present in much lower quantities in our data-sets.

To label the training examples, we use the source position vector in a cartesian coordinate system that is defined by the optical axis of the telescopes, where one coordinate axis is aligned with the horizon. This coordinate system is referred to as the nominal system. The coordinate transformations to the Alt-Az and Ra-Dec systems from the nominal systems, and vice-versa, are done by HAP. 

\begin{figure}
\centering
\includegraphics[width=.46\textwidth]{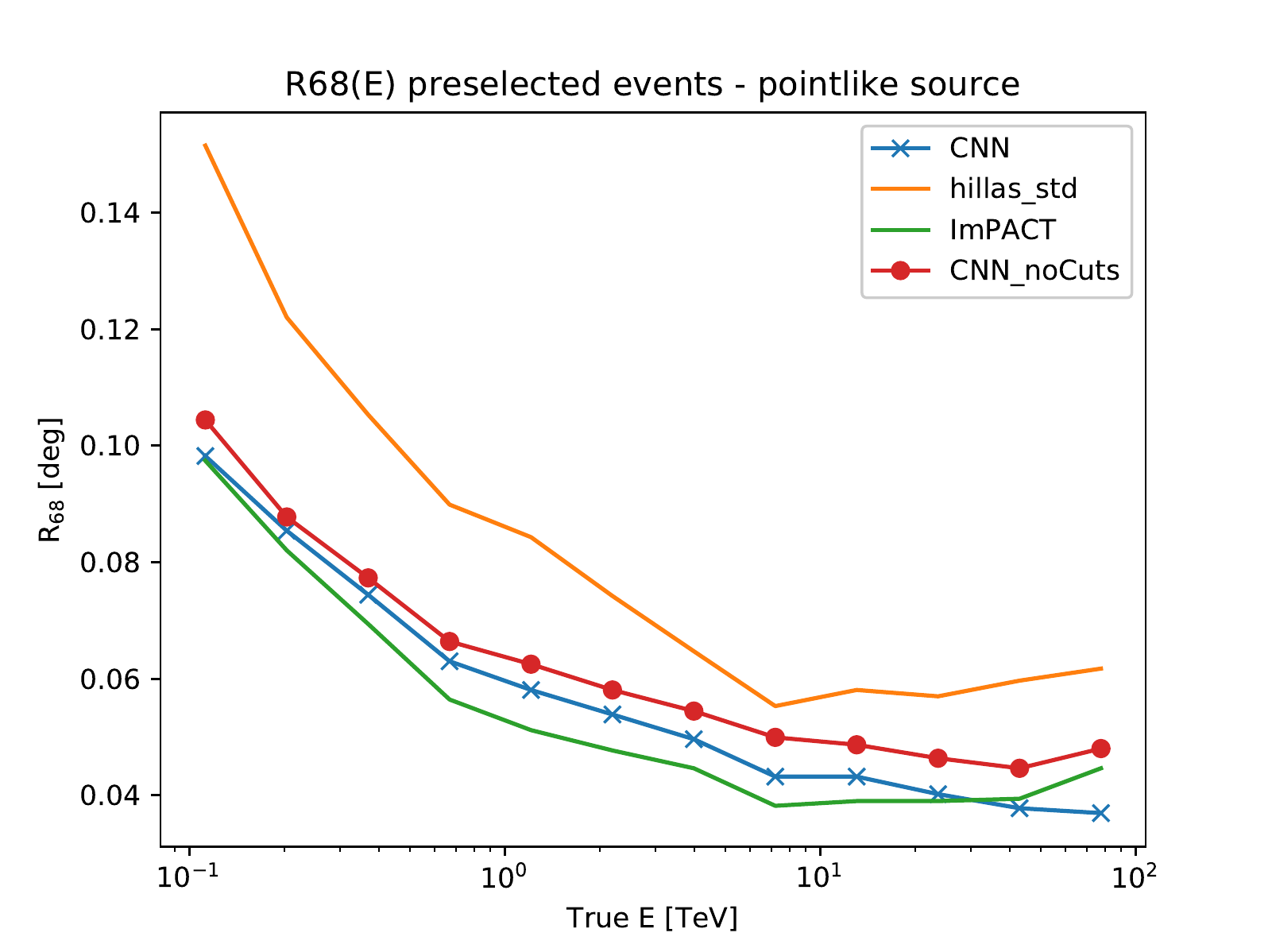}
\caption{Angular resolution vs. true simulated energy at 20$^{\circ}$ zenith angle. The results of two CNN regressors are shown in comparison to the Hillas-based and ImPACT PSFs. The dotted curve refers to a regressor trained without applying pre-selection cuts to the training data, while the 'X'-decorated curve refers to a regressor trained on pre-selected events. All reconstructions are carried out on the same pre-selected benchmark set.}
\label{fig:r68}
\end{figure}

A useful quantitative estimation of the performance of our direction reconstruction on \g-ray simulations is the angular resolution, defined here as the 68\% containment radius (or the 68th percentile) of the reconstructed event positions from a point-like source. This can be taken as a measure of the device PSF. Figure \ref{fig:r68} shows the angular resolution versus true simulated event energy obtained with two CNN-based regressors, Hillas-based reconstruction and ImPACT analysis. The \verb"CNN" label in the legend refers to a regressor that was trained using pre-selected events, while the \verb"CNN_noCuts" label refers to a regressor that was trained without applying pre-selection cuts to the training set. From the figure, both CNN regressors show a significant improvement in the angular resolution, particularly at the low energy range, when compared to the Hillas-based direction reconstruction. Between the regressors, an apparent improvement is achieved by applying pre-selection cuts to the training set (despite the fact that the remaining training set contains only 62\% of the original events). Nevertheless, the \verb"CNN" resolution is still slightly worse than the ImPACT reconstruction. In principle, as our network does not show signs of overfitting with the given architecture, it is also possible that a deeper network will further improve the angular resolution we have shown here.

Despite the fact that the CRNN classifier is not sensitive to relaxation of the pre-selection cuts, the direction regressor shows a similar behavior to the Hillas-based and ImPACT reconstructions with growing values of the local distance, where the angular resolution at energies above 10~TeV grows larger with the increase of the local distance cut. This comes from the fact that the L1 loss is used in the training. As mentioned, such a loss is less sensitive to outliers, which in our case are represented by high energy events, due to the soft energy index of the training set. Using an L2 loss function for the direction reconstruction task generally tends to yield a regressor that performs better at high energies, but has nevertheless worse overall performance than a network learning with an L1 loss. As the real data analysis presented in the following section is done on a source with a very soft spectrum, the choice of the L1 loss seems to be a reasonable one for this work. Nevertheless, to better generlize the regressor one may train on the low energy events using the L1 loss and on the high energy events using the L2 loss.

\section{Source Analysis}
\label{sec:real}

To demonstrate that the results observed with MC simulated events translate well into real-world performance, we test the performance of our CRNN classifier and CNN direction regressor on two source observation sets of the blazar \pks. The first set contains observations of the source in a non-flaring state with the primary goal to test the background rejection performance of the CRNN classifier. We compare these results to those obtained by using a BDT-based background rejection. The second set consists of a single run that was taken during a strong flaring activity of the blazar, an ideal data-set to validate the CNN direction reconstruction performance due to the very low background rate. To conduct the source data analyses, we rely on the \hess~calibration, instrument response functions and lookup tables used for the Hillas and ImPACT~analyses as implemented in the HAP software. Pre-selection cuts, as described in Sec. \ref{sec:bench}, were applied to all reconstruction chains in the following.

\subsection{Observation runs}

\pks~is a bright point-like VHE \g-ray source at redshift $z~=~0.116$~\cite{pks}. For the background rejection test, we selected 14 observation runs which pass the standard H.E.S.S. data quality criteria with a mean observed zenith distance in the range $[19.5^\circ,~20.5^\circ]$, in accordance with the zenith angle of the training set events, from the H.E.S.S. database. The runs were taken between 2004 and 2010 and sum up to a total live-time of 5.9~hours.

The direction reconstruction performance was tested on observations of the \pks~flare that was recorded in 2006 \cite{flare}, when the blazar showed an average flux of $I(>200~\text{GeV}) = (1.72\pm 0.05_{stat}\pm 0.34_{sys})\times 10^{-9}\text{cm}^{-2}\text{s}^{-1}$, corresponding to $\sim7$ times the flux $I(>200 GeV)$ observed from the Crab Nebula (in comparison with the usual 20\% Crab flux of the source). One observation with average zenith angle of $19.2^\circ$ was chosen for this analysis.

\subsection{Analysis}

The separation power of the HAP BDT classifier relies heavily on the Mean Reduced Scaled Width and Length (MRSW and MRSL, respectively) approach introduced in \cite{hess_crab}. These parameters are based on the fact that the width and the length distributions of a \g-ray shower image can be described well with MC simulations, in bins of energy, zenith, azimuth, offset and optical efficiency. The scaled parameters are used as reference to identify \g-like background events that have an elliptical shape but their width and length, do not belong to the same distribution as the \g-ray events with a similar energy. The MRSW and MRSL are a weighted average of the standardized Hillas width and length of the image for all participating telescopes.

A CNN-based neural network is not able to explicitly learn parameters that describe the global distribution of the training set, unless it was specifically trained to predict these parameters. Including all relevant parameters as training labels would require a more complicated and computationally demanding network architecture due to the growing number of parameters needed to find a single function to model the relation between an image to all different labels. Since our labels do not relate to such information, to minimize the loss the network updates the parameters by looking only at intensity gradients in individual images. Thus, an elliptically shaped background image would be classified as signal, even if its MRSW or MRSL are outliers with respect to the image width and length distribution.

To deal with this, we use the HAP lookup tables to obtain the MRSW and MRSL values for each event in the observation runs and use them as additional rejection parameters prior to applying the CRNN classifier. The cut range is defined by the traditional Hillas analysis background rejection scheme, which does not apply a BDT but merely cuts on the MRSW and MRSL values (as defined in \cite{hess_crab}).

This step was not performed in the verification of the CRNN classifier presented in Sec.~\ref{sec:clas} and the quantified estimated performance represents the true classification power of the CRNN classifier (i.e. the only cut used to classify simulated events with the CRNN classifier is the $\zeta$-score cut of 0.5). The reason that the classification MC test shows excellent performance without the additional cuts, stems from the fact that the ratio of signal to background events in the test set is 1:1. The effect of the misclassified \g-like background in the test is seen in the small excess of background counts with high $\zeta$ values in Fig. \ref{fig:z_withcuts}. However, in reality, where the signal to background ratio is at least of the order 1:1000, the influence of \g-like background is much more severe. In the remain of this section, a reference to the CRNN classifier implies a use of the shape cuts~as~well.

\begin{table}[]
\begin{center}
\begin{tabular}{|c|c|c|c|c|}

\hline
Configuration & N$_{\rm{on}}$& $\alpha$N$_{\rm{off}}$ & $\sigma$ & $S/B$ \\
\hline
{\verb"ImPACT_BDT"}   & 704 & 55.8 & 46.1 & 11.6 \\
{\verb"ImPACT_CRNN"} &  832 & 62.3 & 50.8 & 12.4 \\
\hline

\end{tabular}
\end{center}
\caption{Event statistics, significance as calculated by the method of \cite{Li1983} and signal to noise ratio of the 14 runs of \pks~data, for two analyses: an ImPACT direction reconstruction with a BDT classifier (\textit{top row}) and an ImPACT direction reconstruction with a CRNN classifier (\textit{bottom row}).}
\label{tab:pks}
\end{table}

The next step in classifying real world events is an optimization of the $\zeta$-score that will be used to separate events into their appropriate class. Usually, and as done in the case of the HAP~BDT, one looks for an energy independent cut that yields a constant signal efficiency for all energies. We have not optimized our $\zeta$~cut yet in such a way, although we expect an optimized, energy independent value to improve performance. The cut value of 0.9 we chose is based on maximizing the signal to noise ratio in the observation runs of the non flaring \pks, without loosing signal counts compared to the HAP~BDT classifier.

The classifier performance comparison was done by running two separate ImPACT analyses on the 14 \pks~runs: once on events that were flagged as signal by the standard HAP~BDT scheme and once on events that were identified as signal by the CRNN classifier (together with the MRSW and MRSL cuts). The results are summarized in Table \ref{tab:pks}, where the event statistics N$_{\rm{on}}$ (number of events from the source on-region) and $\alpha$N$_{\rm{off}}$ (The estimated background counts) are shown together with the significance $\sigma$ and signal to noise ratio $S/B$. The significance is calculated according to the Li \& Ma method \cite{Li1983}. The table indicates that a CRNN classifier increases both the statistical significance of the source and the ratio of signal events to background events while increasing the number of excess counts (N$_{\rm{on}} - \alpha$N$_{\rm{off}}$). 

The application of our direction reconstruction to a real world source calls for accounting of several correction factors that affect the operation of the telescope in reality. The largest correction is the telescope pointing correction that compensates for small structural, mechanical and thermal deformations. Such considerations are not part of the MC simulations we have used to train our networks. Other factors include atmospheric refraction index and focal length normalization factors. All the system corrections are applied by the HAP software.

\begin{figure}
\centering
\includegraphics[width=.48\textwidth]{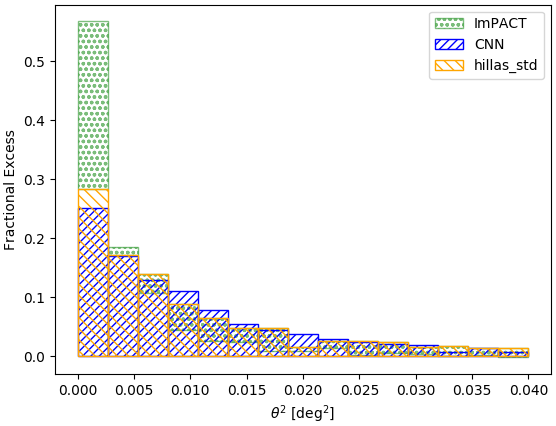}
\caption{The squared angular distance $\theta^2$ distribution for excess events from one flare observation of \pks, using the Hillas, ImPACT and CNN direction reconstruction methods.}
\label{fig:ts}
\end{figure}

The CNN direction reconstruction was done by applying the CRNN classifier to the \pks~flare data-set and reconstructing the direction of the surviving events by means of the CNN regressor that was trained on pre-selected events (see Sec. \ref{sec:dir}). Nevertheless, the real-data performance of the two regressors presented in the previous section is very similar. Figure \ref{fig:ts} shows the distribution of the squared angular distance between the reconstructed event position to the test position $\theta^2$, for excess events from the single \pks~observation in flare state. The $\theta^2$ distribution is plotted for the CNN regressor, a Hillas reconstruction and an ImPACT reconstruction. The Hillas and ImPACT reconstructions use the HAP BDT for background rejection, although the choice of a classifier is less relevant for flare observations. The superior performance of the ImPACT method is clearly evident, where the ImPACT PSF is 0.067\degg, precisely matching the results shown in \cite{impact}. The CNN and Hillas reconstruction are on the same level, with a PSF of 0.102\degg~for both. The PSFs were calculated by using 40 bins over a range of $[0, 0.04]$ in $\theta^2$.

Comparing Figs. \ref{fig:r68} and \ref{fig:ts}, the different performance of the CNN regressor when analyzing simulated events versus real data is clearly evident - particularly when considering the soft spectrum of the source (as the CNN benchmark angular resolution is very similar to the ImPACT angular resolution for low energy events). This is yet another manifestation of the discrepancy between simulated images and real data, which is discussed in Sec. \ref{sec:sum}. 

\begin{figure}
\centering
\includegraphics[width=.48\textwidth]{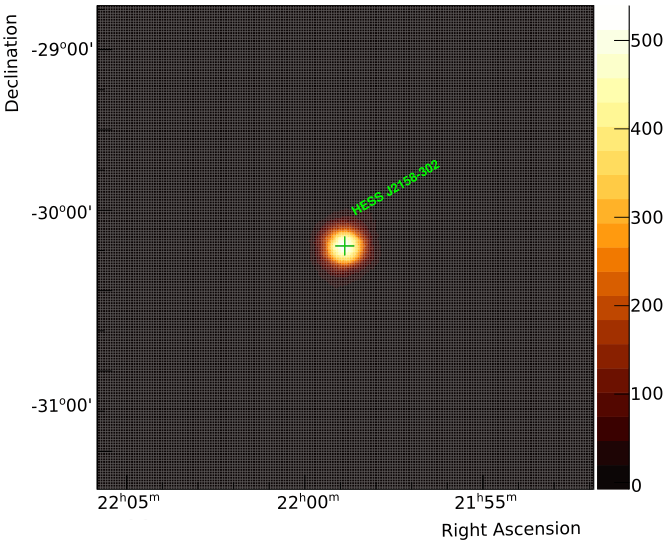}
\caption{A two dimensional distribution of excess events observed in the direction of \pks~for the one flare observation, using the CRNN classifier and CNN regressor.}
\label{fig:sky}
\end{figure}

Fig. \ref{fig:sky} shows a two-dimensional sky map of the excess counts in the direction of \pks. Both figures show that the CNN direction reconstruction is able to detect a point-like source with a similar performance to a Hillas-based analysis. A 2D Gaussian fit of the peak in Fig. \ref{fig:sky} finds a deviation of 11.8~arcsecs from the test position, where the HAP estimated systematic errors are $\sim$30~arcsecs~\cite{hess_crab}.

\section{Summary and Outlook}
\label{sec:sum}

We have demonstrated the improvement in background rejection performance of a convolutional recurrent neural network on a real world bright \g-ray source. Coupled to existing reconstruction techniques, this algorithm seems to have the potential for significant improvement of the analysis of VHE \g-ray sources. Because our classifier is robust against image defects and truncation effects, it could play an even more significant role in the future if a reconstruction technique that is able to deal with truncated images is introduced.

We have also presented a regression model to reconstruct the source sky-coordinates of a \g-ray EAS by applying a deep CNN to IACT images. This algorithm seems to perform reasonably well on simulated images of \g-rays, and can be further improved in the future - even by merely increasing the size of the training set or introducing deeper networks. Applying our direction regressor to real world data seems to be adequate for detecting a point-like source, but needs to be improved to compete with state-of-art methods. Nevertheless, both source analysis results - namely the background rejection and direction reconstruction - are adequate to serve as a proof-of-concept for the viability of a DL based IACT data analysis.

An IACT primary analysis goal that we have not presented here is the energy reconstruction of \g-events. However, it is known that energy can be reconstructed with very high accuracy using the size parameter of the IACT images together with the EAS impact point on the ground. The reconstruction of the ground impact point is done geometrically in a very similar way to reconstruction of the source position, where a finite series of linear transformations separate the two. This implies that the ability to reconstruct the source position is a good indicator of the ability to successfully reconstruct the energy of \g-events with DL techniques.

Energy reconstruction can be carried out in two principle ways. In the first, one would build a network that predicts the ground impact coordinates of the shower and use a lookup table into which the predicted values are input, together with the size parameters. Such an approach can be combined with the direction reconstruction network by adding two more nodes to the output layer. The second way is to build a network that predicts the energy by concatenating a vector that contains the size parameters per event with the vectors of features from a CNN block. We are currently exploring both ways. However, at this stage we merely have a preliminary model which performance cannot be reliably estimated on real world data. A way to quantitatively estimate the energy reconstruction is by measuring a source flux. However, this requires dedicated effective areas that have to be generated by taking into account the cut performance of the background rejection method implemented of the analysis chain. MC simulations to create effective area lookup tables are planned for the near future.

In the future, when larger IACT arrays such as the CTA \cite{cta} are built, one would like to be able to combine observations with different types of telescopes. In the case of \hess~, hybrid reconstruction (i.e. including CT5 data in the event images stack) can be implemented by adding the "new" telescope images to the existing stack of images. With an RNN architecture this is done by simply extending the sequence length. RNNs are flexible enough to deal with different input lengths in the same sequence. With the channel stack, one would need to resample all telescope images to a single and identical square grid. This could introduce the problem of oversampling one image while heavily downsampling another. Of course, memory might become an issue as well as the length of the image stack per event grows.

Another issue with large arrays is the different telescope combinations dedicated for a specific observation-run. Since the number of participating telescopes in a run is certainly smaller than the number of telescopes in the array, an observation-run of a specific source will include a sub-set of the telescopes in the array. Then, a network needs to be trained on the particular telescope combination used for the run. It seems reasonable that run-wise simulations are the way to address this issue.

The disagreements we consistently observe between the performance of the CNN networks on simulated events versus real world events could suggest that a network trained on the complete intensity distribution in the image learns features from simulated images that do not exist in real-data images. This could indicate several difficulties with introducing a DL-based analysis. These difficulties emerge both when such a chain performs a full analysis or even when a single DL-based analysis task is combined with other state-of-the-art methods. In order to calculate fluxes, one relies on MC based effective areas, which are affected by all three analysis tasks. For example, the cut efficiency of the classifier in use directly affects the number of surviving signal events. Since a DL-based classifier acts differently on simulation versus observation data, the effective areas are not reliable when applied to observation events and the derived fluxes could be biased. 

We also note that although a CRNN is seemingly not susceptible to the MC/real-data mismatch, it could certainly suffer from it as well. For example, a network with the CRNN architecture that is trained to distinguish between simulated \emph{proton} images and real-data images becomes astonishingly efficient at performing this task, with an accuracy of 99.5\%. When testing the same classifier on a set comprised of MC \g's (which were not shown to the network during training) and MC protons, it assigns 99.6\% of the events to the 'simulation' class (i.e. the MC protons class in the training set). This illustrates the risk of using simulations for training, as DL methods for computer vision are able to easily find features that do not exist in real-data images. In addition, the performance of the CRNN classifier could improve once the simulated images in the training sets contain similar features to real-data images.

Another issue arises when trying to determine the optimised network architecture for a given task. For example, as mentioned in Sec. \ref{sec:clas}, tuning hyper-parameters to optimize the network performance is done on a hold-out test-set, which is a sample of the training population. However, in our case such test results are not translated well to real world performance because a network that generlizes well on simulated events does not necessarily perform accordingly when applied to observation data.

The MC/real-data discrepancy could indicate an issue with the shape conservation of images resampling, an over-simplification of the telescope response simulation or a strong influence of random noise in real-data images (or a combination of the three). However, our earlier study of resampling methods leads us to rule out the first potential origin of the problem. Due to the importance of this matter we plan to thoroughly study it in the immediate future. For example, the distributions of parameters such as the Hillas parameters, MSRW, MSRL, etc. can be compared to gain more insight into the depth of the problem. Studies on feature importance in a network trained on simulation vs. real-data could assist in this as well. Using e.g. DL based autoencoders on real data images and then applying the decoder part to simulated events could be another possible way to address this issue. We plan to report all these findings in a future publication.

Lastly, in order to be able to properly analyse a real world source, training the network in more zenith bins is required. Dividing the data into energy bins could improve performance as well, but the reduction in training set size will demand a stage of generating simulations in the specific bins, particularly for the higher energies. The run-time of the full analysis depends on the capabilities of the server at hand. Additional GPUs will reduce the run-time considerably. On our two GPU machine the analysis run-times are similar to the ImPACT analysis.

\section*{Acknowledgements}

We thank M. de Naurois, spokesperson for the H.E.S.S. Collaboration, and O. Reimer, chairperson of the Collaboration board, for allowing us to use data from the H.E.S.S. array in this publication. The authors are grateful to C. van~Eldik for many helpful discussions.

\vskip1.6in

\end{document}